\documentclass[aps,twocolumn,showpacs,preprintnumbers,amsmath,amssymb,superscriptaddress,floatfix,nofootinbib]{revtex4}

\usepackage{graphicx}
\usepackage{epsfig}
\usepackage{epstopdf}
\usepackage{hyperref}
\usepackage{amsmath}
\usepackage{amsfonts}
\usepackage{amssymb}
\newcommand{\Slash}[1]{\ooalign{\hfil/\hfil\crcr$#1$}}

\begin{document}

\title{\boldmath Photoproduction of the $f'_2(1525)$ and $K^*_2(1430)$}
\date{\today}

\author{Ju-Jun Xie}
\affiliation{Institute of Modern Physics, Chinese Academy of
Sciences, Lanzhou 730000, China} \affiliation{Research Center for
Hadron and CSR Physics, Institute of Modern Physics of CAS and
Lanzhou University, Lanzhou 730000, China} \affiliation{State Key
Laboratory of Theoretical Physics, Institute of Theoretical Physics,
Chinese Academy of Sciences, Beijing 100190, China}

\author{E.~Oset}
\affiliation{Institute of Modern Physics, Chinese Academy of
Sciences, Lanzhou 730000, China}\affiliation{Departamento de
F\'{\i}sica Te\'orica and IFIC, Centro Mixto Universidad de
Valencia-CSIC Institutos de Investigaci\'on de Paterna, Aptdo.
22085, 46071 Valencia, Spain}

\author{Li-Sheng Geng} \email{lisheng.geng@buaa.edu.cn}
\affiliation{School of Physics and Nuclear Energy Engineering and
International Research Center for Nuclei and Particles in the
Cosmos, Beihang University, Beijing 100191, China}
\affiliation{State Key Laboratory of Theoretical Physics, Institute
of Theoretical Physics, Chinese Academy of Sciences, Beijing 100190,
China}

\begin{abstract}

Assuming  that the $f'_2(1525)$ and $K^*_2(1430)$ resonances are
dynamically generated states from the vector meson-vector meson
interactions in $S$-wave with spin $S=2$, we study the $\gamma p \to
f'_2(1525) p$ and $\gamma p \to K^*_2(1430) \Lambda (\Sigma)$
reactions. These reactions proceed in the following way: the
incoming photon first mutates into a $\rho^0$, $\omega$, or $\phi$
meson via vector meson dominance, which then interacts with the
$\rho^0$, $\omega$ or $K^*$ emitted by the incoming proton to form
the tensor mesons  $f'_2(1525)$ and $K^*_2(1430)$. The picture is
simple and has no free parameters, as all the parameters of the
mechanism have been fixed in previous studies. We predict the
differential and total cross sections of these reactions. The
results can be tested in future experiments and therefore offer new
clues on the nature of these tensor states.

\end{abstract}


\maketitle

\section{Introduction}

Recent observations of the new baryonic  $P_c$
states~\cite{Aaij:2015tga} and the mesonic $XYZ$
states~\cite{Olsen:2014qna,Liu:2013waa,Bodwin:2013nua,Brambilla:2010cs}
by various collaborations have challenged the conventional wisdom
that mesons are made of quark-antiquark pairs and baryons are
composed of three (anti)quarks in the naive quark model. These
findings have attracted a lot of attention from the theory side.
Various explanations of these states have been proposed, such as
molecules, mutiquark compact objects, kinematic effects, or mixtures
of components of different nature. Up to now none of them has been
accepted unanimously. This is not surprising, given limited
experimental constrains and  the fact the various components of a
hadron are not observables themselves. Furthermore, it is quite
likely that a specific reaction or decay process can only review
part of the nature of the hadrons under investigation. Clearly, the
only way to understand the nature of a hadron is to examine it from
all possible ways, both experimentally and theoretically.

Nevertheless, it seems clear that Nature is richer than it is preferred to be.
In this respect, it is not surprising to find out that many low-lying states, even those long believed to be
conventional $q\bar{q}$ (or $qqq$) states, may have large components of other nature. Indeed, it has been shown
that many of the low-lying mesonic states can be understood not only as $q\bar{q}$ states but also as meson-meson molecules,
 dynamically generated in the so-called unitary approaches. One of such examples are
the tensor states: $f_2(1270)$, $f_2'(1525)$, and $K_2^*(1430)$. They  are found dynamically generated
from the vector meson - vector meson interactions~\cite{Molina:2008jw,Geng:2008gx,Geng:2009gb},
obtained in the coupled-channel Bethe-Salpeter equations by unitarizing the tree-level hidden gauge diagrams ~\cite{hidden1,hidden2,hidden3,hidden4}.

The molecular nature of these tensor states has been extensively
tested in a large number of processes, for instance,  the two-photon
decay of the $f_2(1270)$~\cite{yamagata}; the two-photon and one
photon-one vector decays of the $f_2(1270)$, $f'_2(1525)$ and
$K^*_2(1430)$~\cite{Branz:2009cv};
 the $J/\psi \to \phi(\omega)
f_2(1270), ~f'_2(1525)$ and $J/\psi \to K^{*0}(892)
\bar{K}^{*0}_2(1430)$ decays~\cite{alberzou};  the radiative decay of $J/\psi$ into
$f_2(1270)$ and $f'_2(1525)$ ~\cite{Geng:2009iw};
the $\psi (2S)$ decays into $\omega(\phi) f_2(1270)$, $\omega(\phi)
f'_2(1525)$, $K^{*0}(892) \bar{K}^{*\,0}_2(1430)$ and the radiative
decays of $\Upsilon (1S),\Upsilon (2S), \psi (2S)$ into $\gamma
f_2(1270)$, $\gamma f'_2(1525)$, $\gamma f_0(1370)$, and $\gamma
f_0(1710)$ ~\cite{dai2s,Dai:2015cwa};
 the ratio of the decay widths of
$\bar B^0_s \to J/\psi f_2(1270)$ to $\bar B^0_s \to J/\psi
f'_2(1525)$ ~\cite{xiebdec}. The agreement with experimental data turns out to be quite good in general, providing
support to the  underlying assumption that these states contain large meson-meson components.

In a recent work~\cite{Xie:2014twa}, taking the molecular picture
for the $f_2(1270)$ resonance, the $\gamma p \to p f_2(1270)$
reaction has been studied. It was found that the theoretical results
of the differential cross sections are in agreement with the
experimental data of Ref.~\cite{Battaglieri:2009aa}, providing first
support for the molecular picture of the $f_2(1270)$ state in a
baryonic reaction. In this work, we extend the formalism proposed in
Ref.~\cite{Xie:2014twa} to study the $\gamma p \to f'_2(1525) p$ and
$\gamma p \to K^*_2(1430) \Lambda (\Sigma)$ reactions. One should
stress again that as the only way to unravel the nature of a hadron
is via different reactions and decay processes, we deem such studies
very timely and important.

The present article is organized as follows. In
Sec.~\ref{sec:formalism}, we introduce the formalism and the main
ingredients of the model. In Sec.~\ref{sec:results}, we present our
main results, and a short summary and conclusions are given
in Sec.~\ref{sec:summary}.

\section{Formalism and ingredients} \label{sec:formalism}

\subsection{Feynman amplitudes}

\begin{figure}[htbp]
\begin{center}
\includegraphics[scale=1.2]{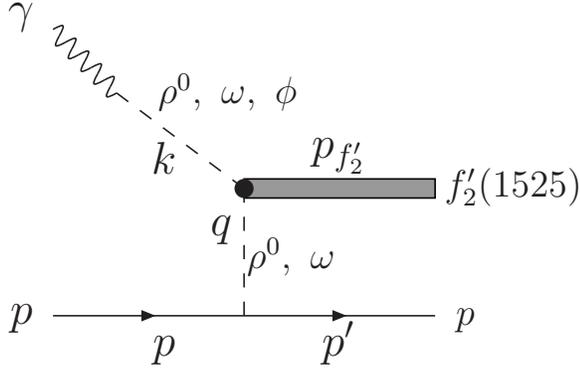}
\caption{Diagrammatic representation of the $f'_2(1525)$
photoproduction, where $k,~p,~p',q,p_{f_2}$ are the four-momentum of
the involved particles and $q=p'-p$.} \label{Fig:gammaptof2p}
\end{center}
\end{figure}

\begin{figure}[htbp]
\begin{center}
\includegraphics[scale=1.]{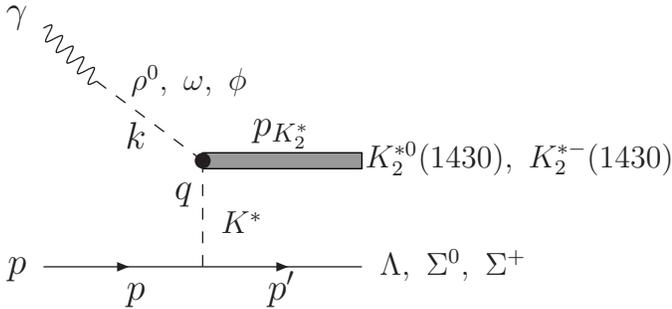}
\caption{Diagrammatic representation of the $\gamma p \to
K^*_2(1430)\Lambda (\Sigma)$ reaction.} \label{Fig:gammaptok2star}
\end{center}
\end{figure}

From the perspective that the $f'_2(1525)$ and $K^*_2(1430)$
resonances are dynamically generated from the vector-vector ($VV$)
interactions, the $f'_2(1525)$ and $K^*_2(1430)$ photoproductions
proceed via the creation of two vector mesons by the $\gamma p$
initial state in a primary step and the following interaction
(rescattering) of the two vector mesons, thus dynamically generating
the resonance. This corresponding Feynman diagrams are shown in
Fig.~\ref{Fig:gammaptof2p} for the $\gamma p \to f'_2(1525) p$
reaction and Fig.~\ref{Fig:gammaptok2star} for the $\gamma p \to
K^*_2(1430)\Lambda (\Sigma)$ reaction.

\begin{figure}[htbp]
\begin{center}
\includegraphics[scale=0.7]{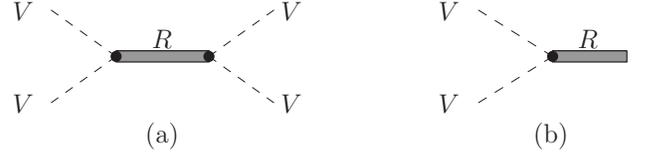}
\caption{(a): The $VV \to VV$ amplitude dominated by the
$f'_2(1525)$ or $K^*_2(1430)$ pole; (b): Representation of the
$f'_2(1525)$ or $K^*_2(1430)$ coupling to $VV$.}
\label{Fig:rhorhotof2}
\end{center}
\end{figure}

As can be seen from Figs.~\ref{Fig:gammaptof2p} and
\ref{Fig:gammaptok2star}, the photon first gets converted into one
vector meson, a characteristic of the local hidden gauge formalism,
which then interacts with the other vector emitted by the proton. To
evaluate the Feynman amplitudes, we need the coupling of the tensor
meson to the respective vector mesons, $g_{T}^{VV}$, the
$\gamma$-$V$ coupling, and the $V NN$ coupling. In the unitary
approach, the amplitude close to a pole that represents a resonance
can be written in the following way
\begin{eqnarray}
t_{\rm pole} &\simeq& \frac{\left(g^{VV}_T \right)^2 P^{(2)}_{\rm initial} P^{(2)}_{\rm
final}}{s - s_R},       \label{tpole} \\
P^{(2)}_{\rm initial} &=& \frac{1}{2}(\epsilon^{(1)}_i
\epsilon^{(2)}_j + \epsilon^{(1)}_j \epsilon^{(2)}_i)
-\frac{1}{3}\epsilon^{(1)}_l \epsilon^{(2)}_l \delta_{ij}, \\
P^{(2)}_{\rm final} &=& \frac{1}{2}(\epsilon^{(3)}_i
\epsilon^{(4)}_j + \epsilon^{(3)}_j \epsilon^{(4)}_i)
-\frac{1}{3}\epsilon^{(3)}_m \epsilon^{(4)}_m \delta_{ij},
\end{eqnarray}
where $s_R$ is the pole position and $g^{VV}_T$ the coupling of the
resonance to the $VV$ component in isospin $I=0(1/2)$ and spin
$S=2$. The Eq.~(\ref{tpole}) is the representation of a resonance
amplitude, for instance the $f'_2(1525)$ and $K^*_2(1430)$ in the
present case, as shown in Fig.~\ref{Fig:rhorhotof2} (a). The
$P^{(2)}_\mathrm{initial/final}$  projects the initial and final
$VV$ pair into spin two.  Then the coupling of a tensor resonance to
$VV$ is given by the diagram of Fig.~\ref{Fig:rhorhotof2} (b), and
is expressed in terms of the following vertex~\cite{yamagata}
\begin{eqnarray}
t_{R \to VV} = g^{VV}_T P^{(2)}_{\rm initial} , \label{tRrhorho}
\end{eqnarray}
where the values for $g^{VV}_T$ are shown in Table~\ref{Tab:tvvcoupling}, taken from
Ref.~\cite{Geng:2008gx}.

\begin{table}[htbp]
\caption{The coupling constants of $f'_2(1525)$ and $K^*_2(1430)$ to
$VV$.}
\begin{center}
\begin{tabular}{ccc}
\hline\hline
Resonance & Channel & $g^{VV}_T$ (MeV)  \\
\hline
$f'_2(1525)$ & $\rho \rho$  & $(-2443,i649)$  \\
             & $\omega \omega$  & $(-2709,i8)$  \\
             & $ \phi\omega$  & $(5016,-i17)$  \\
$K^*_2(1430)$ & $\rho K^*$ & $(10901,-i71)$   \\
& $\omega K^*$ & $(2267,-i13)$   \\
& $\phi K^*$ & $(-2898,i17)$   \\
\hline \hline
\end{tabular}
\end{center} \label{Tab:tvvcoupling}
\end{table}

The $\gamma$-$V$ conversion vertex can be obtained from the local
hidden gauge Lagrangians~\cite{hidden1,hidden2,hidden3,hidden4} (see
Ref.~\cite{Nagahiro:2008cv} for a practical set of rules) and one
has~\cite{angelsphotok}
\begin{eqnarray}
-i t_{\gamma V} = -i C_{\gamma V} \frac{eM^2_V}{g} \epsilon_{\mu}(V)
\epsilon^{\mu}(\gamma), \label{trhogamma}
\end{eqnarray}
with
\begin{eqnarray}
g = \frac{M_{\rho}}{2f};~~~~f=93~~{\rm
MeV};~~~~\frac{e^2}{4\pi}=\frac{1}{137},
\end{eqnarray}
and
\begin{eqnarray}
C_{\gamma \rho} = \frac{1}{\sqrt{2}};~~~~C_{\gamma \omega} =
\frac{1}{3\sqrt{2}};~~~~C_{\gamma \phi} = -\frac{1}{3}.
\end{eqnarray}

The other ingredient that we need is the vector-baryon-baryon
vertex, which is given by the Lagrangian
\begin{eqnarray}
{\cal L}_{BBV} = g (<\bar{B}\gamma^{\mu}[V_{\mu},B]> +
<\bar{B}\gamma^{\mu}B><V_{\mu}>).\label{lbbv}
\end{eqnarray}
From this, one can easily obtain the $\rho^0 pp$, $\omega pp$, $K^{*+} p \Lambda$,
$K^{*+} p \Sigma^0$, and $K^{*0} p \Sigma^+$ vertices,
\begin{eqnarray}
-it_{\rho^0 pp} &=& i \frac{g}{\sqrt{2}} \bar{p}\gamma^{\mu} p
\epsilon_{\mu}(\rho^0), \label{rhopp} \\
-it_{\omega pp} &=& i \frac{3g}{\sqrt{2}} \bar{p}\gamma^{\mu} p
\epsilon_{\mu}(\omega),  \label{omepp} \\
-it_{K^{*+} p \Lambda} &=& -i \frac{3g}{\sqrt{6}}
\bar{\Lambda}\gamma^{\mu} p
\epsilon_{\mu}(K^*),  \label{kstarplambda} \\
-it_{K^{*+} p \Sigma^0} &=& -i \frac{g}{\sqrt{2}}
\bar{\Sigma}\gamma^{\mu} p
\epsilon_{\mu}(K^*), \label{kstarpsigmazero} \\
-it_{K^{*0} p \Sigma^+} &=& -i g \bar{\Sigma}\gamma^{\mu} p
\epsilon_{\mu}(K^*). \label{kstarpsigmaplus}
\end{eqnarray}

There is one more subtlety  to consider. The amplitude of
Eq.~(\ref{tpole}) is evaluated for a $VV$ state in the unitary
normalization, which for $I=0$ and $I = 1/2$ are given as follows (recall
$|\rho^+> = -|1,+1>$),
\begin{eqnarray}
&& |\rho\rho,I=0> = -\frac{1}{\sqrt{6}} (\rho^+ \rho^- + \rho^-
\rho^+ + \rho^0 \rho^0), \label{f2rhorho} \\
&& |\omega\omega,I=0> = \frac{1}{\sqrt{2}} (\omega \omega), \label{f2omeome} \\
&& |\phi\omega,I=0> = (\phi \omega), \\
&& |\rho K^*,I=1/2,I_3 = 1/2> = \sqrt{\frac{2}{3}}(\rho^+ K^{*0}) ,\nonumber \\
&& - \sqrt{\frac{1}{3}}(\rho^0 K^{*+}), \\
&& |\rho K^*,I=1/2,I_3 = -1/2> = \sqrt{\frac{1}{3}}(\rho^0 K^{*0}),
\nonumber \\
&& - \sqrt{\frac{2}{3}}(\rho^- K^{*+}), \\
&& |\omega K^*,I=1/2> = (\omega K^*), \\
&& |\phi K^*,I=1/2> = (\phi K^*).
\end{eqnarray}

Considering the vertices described above, we obtain the weights $W^{VV}_T$ for
$\gamma p \to f'_2(1525) p$ and $\gamma p \to K^*_2(1430)\Lambda
(\Sigma)$ reactions (Table~\ref{Tab:weights}). They account for the
factor of Eq.~(\ref{trhogamma}),
Eqs.~(\ref{rhopp}-\ref{kstarpsigmaplus}), the components of the $VV$
state in the good normalization and the couplings of the resonances
to the $VV$ channel. In the case of two identical particles the
coupling is multiplied by an extra factor of $\sqrt{2}$ to restore the
good normalization from the couplings calculated in
Ref.~\cite{Geng:2008gx} in the unitary approach as shown in Eqs.~(14-20).

\begin{table}[htbp]
\caption{Weights for the $\gamma p \to f'_2(1525) p$ and $\gamma p
\to K^*_2(1430)\Lambda (\Sigma)$ reactions. The  $f'_2(1525)$ and $K^*_2(1430)$ are
abbreviated as $f'_2$ and $K^*_2$, respectively.}
\begin{center}
\begin{tabular}{ccc}
\hline\hline
Reaction &   & $W^{VV}_T$  \\
\hline
$\gamma p \to f'_2(1525) p$ & $\rho^0 \rho^0$  & $-\frac{e}{\sqrt{6}} g^{\rho \rho}_{f'_2}$  \\
             & $\omega \omega$  & $\frac{e}{\sqrt{2}} g^{\omega \omega}_{f'_2}$  \\
             & $ \phi\omega$  & $-\frac{e}{\sqrt{2}} g^{\phi \omega }_{f'_2}$  \\
$\gamma p \to K^{*+}_2(1430) \Lambda$ & $\rho^0 K^{*+}$ & $\frac{e}{2} g^{\rho K^*}_{K^*_2}$   \\
& $\omega K^{*+}$ &  $-\frac{e}{2\sqrt{3}} g^{\omega K^*}_{K^*_2}$   \\
& $\phi K^{*+}$ & $\frac{e}{\sqrt{6}}  g^{\phi K^*}_{K^*_2}$   \\
$\gamma p \to K^{*+}_2(1430) \Sigma^0$ & $\rho^0 K^{*+}$ & $\frac{e}{2\sqrt{3}} g^{\rho K^*}_{K^*_2}$   \\
& $\omega K^{*+}$ &  $-\frac{e}{6} g^{\omega K^*}_{K^*_2}$   \\
& $\phi K^{*+}$ & $\frac{e}{3\sqrt{2}}  g^{\phi K^*}_{K^*_2}$   \\
$\gamma p \to K^{*0}_2(1430) \Sigma^+$ & $\rho^0 K^{*0}$ & $-\frac{e}{\sqrt{6}} g^{\rho K^*}_{K^*_2}$   \\
& $\omega K^{*0}$ &  $-\frac{e}{3\sqrt{2}} g^{\omega K^*}_{K^*_2}$   \\
& $\phi K^{*0}$ & $\frac{e}{3} g^{\phi K^*}_{K^*_2}$   \\
\hline \hline
\end{tabular}
\end{center} \label{Tab:weights}
\end{table}

Gauge invariance imposes a stringent constraint on photonuclear
processes, although sometimes not all of the terms needed to have
gauge invariance are numerically
relevant~\cite{Borasoy:2005zg,Doring:2005bx}. Nevertheless, in the
present case, a thorough study of gauge invariance was conducted in
Ref.~\cite{Nagahiro:2008cv} for the radiative decay of axial vector
mesons within the local hidden gauge approach, and in particular in
Ref.~\cite{yamagata} for the amplitude $\rho \rho \to \rho \gamma$,
which is similar to what we have here, with the two vector mesons
interacting to produce the tensor states. There it is concluded that
gauge invariance is encoded in the effective coupling of the tensor
states to the two vector mesons.

Considering the weights given above, the $T$ matrix for the diagram
of Fig.~\ref{Fig:gammaptof2p} is given by
\begin{eqnarray}
&& -i T_{\gamma p \to f'_2(1525)p} = -i e(-\frac{g^{\rho
\rho}_{f'_2}}{\sqrt{6}} + \frac{g^{\omega \omega}_{f'_2}}{\sqrt{2}}
- \frac{g^{ \phi \omega}_{f'_2}}{\sqrt{2}}) \nonumber \\
&& \{ \frac{1}{2} [ \epsilon_i(\gamma) \epsilon_j(V) +
\epsilon_j(\gamma) \epsilon_i(V) ] -\frac{1}{3}\epsilon_m(\gamma)
\epsilon_m(V) \delta_{ij} \} \nonumber \\
&& \frac{1}{q^2 - m^2_{V}}
<p(M')|\gamma^{\mu}\epsilon_{\mu}(V)|p(M)>, \label{tgammap1}
\end{eqnarray}
with $M$ and $M'$ the  third spin component of the initial and final
proton. The $V$ stands for the exchanged $\rho^0$ or $\omega$. We
take $m_V = m_{\rho} = m_{\omega} = 780$ MeV in the present
calculation. Next, we perform the sum over the polarizations of
the vector meson exchanged in Fig.~\ref{Fig:gammaptof2p} and then we
obtain
\begin{eqnarray}
&& T_{\gamma p \to f'_2(1525)p} = e(-\frac{g^{\rho
\rho}_{f'_2}}{\sqrt{6}} + \frac{g^{\omega \omega}_{f'_2}}{\sqrt{2}}
- \frac{g^{ \phi \omega}_{f'_2}}{\sqrt{2}}) \frac{1}{q^2 - m^2_V}
\nonumber \\
&& [\frac{1}{2}\epsilon_i(\gamma) (-g_{j\mu} +
\frac{q_jq_{\mu}}{m^2_V})  + \frac{1}{2} \epsilon_j(\gamma) (-g_{i \mu} + \frac{q_i q_{\mu}}{m^2_V}) \nonumber \\
&&  -\frac{1}{3}\epsilon_m(\gamma) \delta_{ij} (-g_{m \mu} +
\frac{q_m q_{\mu}}{m^2_V})] \nonumber \\
&&  <p(M')|\gamma^{\mu}|p(M)>. \label{tgammap2}
\end{eqnarray}

Following the same procedure, we obtain the transition amplitudes
for $\gamma p\to K^{*}_2(1430)\Lambda(\Sigma)$:
\begin{eqnarray}
&& T_{\gamma p \to K^{*+}_2(1430) \Lambda} = e(\frac{g^{\rho
K^*}_{K^*_2}}{2} - \frac{g^{\omega K^*}_{K^*_2}}{2\sqrt{3}} +
\frac{g^{ \phi K^*}_{K^*_2}}{\sqrt{6}}) \nonumber \\
&& \frac{1}{q^2 - m^2_{K^*}} [\frac{1}{2}\epsilon_i(\gamma)
(-g_{j\mu} +
\frac{q_jq_{\mu}}{m^2_{K^*}}) \nonumber \\
&& + \frac{1}{2} \epsilon_j(\gamma) (-g_{i \mu} + \frac{q_i
q_{\mu}}{m^2_{K^*}})  -\frac{1}{3}\epsilon_m(\gamma) \delta_{ij}
(-g_{m \mu} +
\frac{q_m q_{\mu}}{m^2_{K^*}})] \nonumber \\
&& \times  <\Lambda(M')|\gamma^{\mu}|p(M)>, \\
&& T_{\gamma p \to K^{*+}_2(1430) \Sigma^0} = e(\frac{g^{\rho
K^*}_{K^*_2}}{2\sqrt{3}} - \frac{g^{\omega K^*}_{K^*_2}}{6} +
\frac{g^{ \phi K^*}_{K^*_2}}{3\sqrt{2}}) \nonumber \\
&&  \frac{1}{q^2 - m^2_{K^*}} [\frac{1}{2}\epsilon_i(\gamma)
(-g_{j\mu} +
\frac{q_jq_{\mu}}{m^2_{K^*}}) \nonumber \\
&&  + \frac{1}{2} \epsilon_j(\gamma) (-g_{i \mu} + \frac{q_i
q_{\mu}}{m^2_{K^*}}) -\frac{1}{3}\epsilon_m(\gamma) \delta_{ij}
(-g_{m \mu} +
\frac{q_m q_{\mu}}{m^2_{K^*}})] \nonumber \\
&& \times  <\Sigma(M')|\gamma^{\mu}|p(M)>, \\
&& T_{\gamma p \to K^{*0}_2(1430) \Sigma^+} = e(-\frac{g^{\rho
K^*}_{K^*_2}}{\sqrt{6}} - \frac{g^{\omega K^*}_{K^*_2}}{3\sqrt{2}} +
\frac{g^{ \phi K^*}_{K^*_2}}{3}) \nonumber \\
&&  \frac{1}{q^2 - m^2_{K^*}} [\frac{1}{2}\epsilon_i(\gamma)
(-g_{j\mu} +
\frac{q_jq_{\mu}}{m^2_{K^*}}) \nonumber \\
&&  + \frac{1}{2} \epsilon_j(\gamma) (-g_{i \mu} + \frac{q_i
q_{\mu}}{m^2_{K^*}})  -\frac{1}{3}\epsilon_m(\gamma) \delta_{ij}
(-g_{m \mu} + \frac{q_m q_{\mu}}{m^2_{K^*}})] \nonumber \\
&& \times <\Sigma(M')|\gamma^{\mu}|p(M)>, \label{tgammaK2}
\end{eqnarray}
where we take $m_{K^*} = m_{K^{*+}} = m_{K^{*0}} = 893.1$ MeV.

In Eqs.~(\ref{tgammap2})-(\ref{tgammaK2}),  the latin
indices run over $1$, $2$, $3$ and the $\mu$ index from $0, 1, 2,
3$.

Then one can easily calculate $\bar{\sum}\sum|T|^2$. Here we give
explicitly the case of the $\gamma p \to f'_2(1525) p$ reaction, as an
example,
\begin{eqnarray}
&& \bar{\sum}\sum|T|^2 = \frac{e^2}{96 m^2_p (q^2 -
m^2_V)^2}|-\frac{g^{\rho \rho}_{f'_2}}{\sqrt{6}} + \frac{g^{\omega
\omega}_{f'_2}}{\sqrt{2}} - \frac{g^{ \phi
\omega}_{f'_2}}{\sqrt{2}}|^2 \nonumber \\
&& \sum_{\gamma~{\rm pol.}}
\sum_{i,j,m,l}\sum_{\mu,\mu'}[\frac{1}{2}\epsilon_i(\gamma)
(-g_{j\mu} + \frac{q_jq_{\mu}}{m^2_V}) \nonumber \\
&& + \frac{1}{2} \epsilon_j(\gamma) (-g_{i \mu} + \frac{q_i
q_{\mu}}{m^2_V})
 -\frac{1}{3}\epsilon_m(\gamma) \delta_{ij} (-g_{m \mu} +
\frac{q_m q_{\mu}}{m^2_V})] \nonumber \\
&& [\frac{1}{2}\epsilon_i(\gamma) (-g_{j\mu'} +
\frac{q_jq_{\mu'}}{m^2_V})   + \frac{1}{2} \epsilon_j(\gamma) (-g_{i
\mu'} + \frac{q_i q_{\mu'}}{m^2_V}) \nonumber \\
&&  -\frac{1}{3}\epsilon_l(\gamma) \delta_{ij}
(-g_{l \mu'} + \frac{q_l q_{\mu'}}{m^2_V})] \nonumber \\
&& Tr[(\Slash{p'} + m_p)\gamma^{\mu}(\Slash{p} + m_p)\gamma^{\mu'}],
\end{eqnarray}
where  all the indices and the two photon
polarizations should be summed over, with the following expressions of the latter,
\begin{eqnarray}
\epsilon^{(1)}(\gamma) = \left (
\begin{array}{c}
1   \\
0   \\
0
\end{array}
\right ); ~~~~\epsilon^{(2)}(\gamma) = \left ( \begin{array}{c}
0   \\
1   \\
0
\end{array} \right ),
\end{eqnarray}
where we have assumed that the photon travels in the $Z$ direction.

\subsection{Differential cross section}

The differential cross section for $\gamma p \to f'_2(1525) p$ and
$\gamma p \to K^*_2(1430) \Lambda (\Sigma)$ reactions are given by
\begin{eqnarray}
\frac{d\sigma}{dt} = \frac{m_i^2}{16\pi s |\vec{k}|^2}
\bar{\sum}\sum |T|^2, \label{dsigmadt}
\end{eqnarray}
with $s$ the invariant mass squared of the $\gamma p$ system, and
$m^2_i = m^2_p$ for $\gamma p \to f'_2(1525) p$ reaction, $m^2_i =
m_p m_{\Lambda}$ for $\gamma p \to K^{*+}_2(1430) \Lambda$ reaction,
and $m^2_i = m_p m_{\Sigma}$ for $\gamma p \to K^*_2(1430) \Sigma$
reaction.~\footnote{We take $m_{\Sigma^0} = m_{\Sigma^+} = 1191$ MeV
and $m_{K_2^{*+}} = m_{K_2^{*0}} = 1429$ MeV in this work.} The
$\vec{k}$ is the three momenta of the initial photon in the center
of mass frame (c.m.), and $t= q^2 = (p-p')^2$.

The Eq.~(\ref{dsigmadt}) can be generalized for the case when the
$f'_2(1525)$ ($K^*_2(1430)$) is explicitly allowed to decay into
$K\bar{K}$ ($K\pi$) by working out the three body phase space and we
find
\begin{eqnarray}
\frac{d^2\sigma}{dM_{\rm inv} dt} &=& \frac{m^2_i}{8\pi^2 s
|\vec{k}|^2} \frac{M^2_{\rm inv} \Gamma_{i}}{|M^2_{\rm inv} -
M^2_{R} + i
M_{\rm inv} \Gamma_{R}|^2} \nonumber \\
&& \times \bar{\sum}\sum |T|^2, \label{dsigmadmdt}
\end{eqnarray}
where $M_{\rm inv}$ is the invariant mass distribution of the
$K\bar{K}$ or $K\pi$, $\Gamma_{R}$ is the total decay width of the
$f'_2(1525)$ or $K^*_2(1430)$ and $\Gamma_{i}$ is the partial decay
width of the $f'_2(1525) \to K \bar{K}$ or $K^*_2(1430) \to K \pi$.
In the present study, we choose the following decay modes: $f'_2(1525) \to K^+ K^-$,
$K^{*+}_2(1430) \to K^0 \pi^+$, and $K^{*0}_2(1430) \to K^+ \pi^-$.
The $f'_2(1525) \to K^+ K^-$ decay accounts for $1/2$ of the
$K\bar{K}$ decay of the $f'_2(1525)$ which is $89\%$ of the
$\Gamma_{f'_2(1525)}$, while the $K^{*+}_2(1430) \to K^0 \pi^+$ or
$K^{*0}_2(1430) \to K^+ \pi^-$ decay accounts for $2/3$ of the
$K\pi$ decay of the $K^*_2(1430)$ which is $50\%$ of
$\Gamma_{K^*_2(1430)}$. Since the $f'_2(1525) \to K \bar{K}$ and
$K^*_2(1430) \to K \pi$ decays are in $D$-wave, in order to have
$\Gamma_i$ and $\Gamma_R$ in the range of invariant masses that we
consider, we take
\begin{eqnarray}
&& \Gamma_{f'_2 \to K \bar{K}}(M_{\rm inv}) = \Gamma_{K
\bar{K}}^{\rm on} (\frac{\tilde{q}_{K\bar{K}}}{\bar{q}_{K\bar{K}}} )^5 \frac{M^2_{f'_2}}{M^2_{\rm inv}} , \\
&& \Gamma_{f'_2} (M_{\rm inv}) = 0.89 \Gamma_{f'_2}^{\rm on} (
\frac{\tilde{q}_{K\bar{K}}}{\bar{q}_{K\bar{K}}} )^5
\frac{M^2_{f'_2}}{M^2_{\rm inv}} + 0.11 \Gamma_{f'_2}^{\rm on},
\end{eqnarray}
with $\Gamma^{\rm on}_{f'_2} = 73$ MeV, $\Gamma^{\rm on}_{K\bar{K}}
= 32.5$ MeV, $M_{f'_2} = 1525$ MeV~\cite{Agashe:2014kda}, and
\begin{eqnarray}
\tilde{q}_{K\bar{K}} &=& \frac{\lambda^{1/2}(M^2_{\rm inv}, m^2_{K},
m^2_{\bar{K}})}{2M_{\rm inv}}, \\
\bar{q}_{K\bar{K}} &=& \frac{\lambda^{1/2}(M^2_{f'_2}, m^2_{K},
m^2_{\bar{K}})}{2M_{f'_2}},
\end{eqnarray}
where $\lambda$ is the K\"allen function with $\lambda(x,y,z) =
(x-y-z)^2 - 4yz$.

Similarly, for the $K^*_2$ decay modes, we take
\begin{eqnarray}
&& \Gamma_{K^*_2 \to K \pi}(M_{\rm inv}) = \Gamma_{K
\pi}^{\rm on} (\frac{\tilde{q}_{K\pi}}{\bar{q}_{K\pi}} )^5 \frac{M^2_{K^*_2}}{M^2_{\rm inv}} , \\
&& \Gamma_{K^*_2} (M_{\rm inv}) = 0.5 \Gamma_{K^*_2}^{\rm on} (
\frac{\tilde{q}_{K\bar{K}}}{\bar{q}_{K\bar{K}}} )^5
\frac{M^2_{K^*_2}}{M^2_{\rm inv}} + 0.5 \Gamma_{K^*_2}^{\rm on},
\end{eqnarray}
with $\Gamma^{\rm on}_{K^*_2} = 104$ MeV, $\Gamma^{\rm on}_{K\pi} =
34.7$ MeV~\cite{Agashe:2014kda}, and
\begin{eqnarray}
\tilde{q}_{K\pi} &=& \frac{\lambda^{1/2}(M^2_{\rm inv}, m^2_{K},
m^2_{\pi})}{2M_{\rm inv}}, \\
\bar{q}_{K\pi} &=& \frac{\lambda^{1/2}(M^2_{K^*_2}, m^2_{K},
m^2_{\pi})}{2M_{K^*_2}}.
\end{eqnarray}

\section{Numerical results} \label{sec:results}

In Ref.~\cite{Xie:2014twa} three models were considered, one is the
one we exposed here, and the other two contained an additional tensor
$\rho NN$ coupling and Regge propagators. The results
obtained there were very similar and in this exploratory work we
perform calculations only with one of the models as specified above.

\begin{figure}[htbp]
\begin{center}
\includegraphics[scale=0.45]{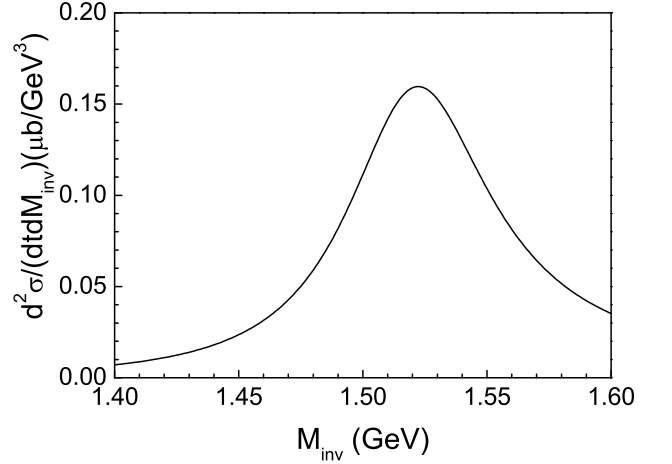}
\caption{Theoretical predictions for the $D$-wave $K \bar{K}$ mass
distribution of the $\gamma p \to p K^+ K^-$ reaction at $E_{\gamma} =
3.4$ GeV and $t = -1.2$ GeV$^2$} \label{Fig:dsigdmdtf2p}
\end{center}
\end{figure}

In Fig.~\ref{Fig:dsigdmdtf2p} we show $\frac{d^2\sigma}{dM_{\rm
inv}dt}$  for the $\gamma p \to p K^+K^-$ reaction for at
$E_{\gamma} = 3.4$ GeV and $t = -1.2$ GeV$^2$, where $M_{\rm inv}$
is the invariant mass of $K^+K^-$ system. The $f_2'(1525)$ resonance
is clearly seen, peaking around $M_{\rm inv} = 1525$ MeV with an
apparent width of about 60 MeV.

\begin{figure}[htbp]
\begin{center}
\includegraphics[scale=0.45]{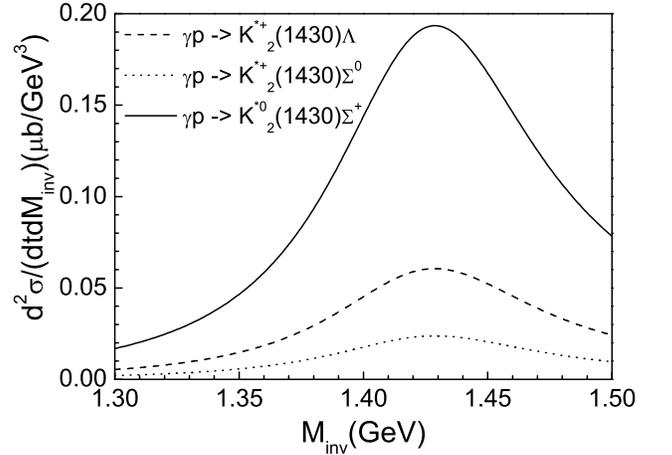}
\caption{Theoretical predictions for the $D$-wave $K \pi$ mass
distribution of the $\gamma p \to \Lambda (\Sigma) K \pi$ reaction at
$E_{\gamma} = 3.4$ GeV and $t = -1.2$ GeV$^2$}
\label{Fig:dsigdmdtk2star}
\end{center}
\end{figure}

In Fig.~\ref{Fig:dsigdmdtk2star} we show  $\frac{d^2\sigma}{dM_{\rm
inv}dt}$  for the $\gamma p
\to \Lambda (\Sigma) K \pi$ reaction at  $E_{\gamma} = 3.4$ GeV and $t = -1.2$ GeV$^2$,
where $M_{\rm inv}$ is the invariant mass of the $K
\pi$ system.  The $K^*_2(1430)$ is clearly seen in all the three decay modes, but the magnitude of the decay modes
are quite different. Experimental confirmation of such a hierarchy constitutes a valid check of the molecular picture
of the $K_2^*(1430)$ and the reaction mechanism advocated here.

\begin{figure}[htbp]
\begin{center}
\includegraphics[scale=0.45]{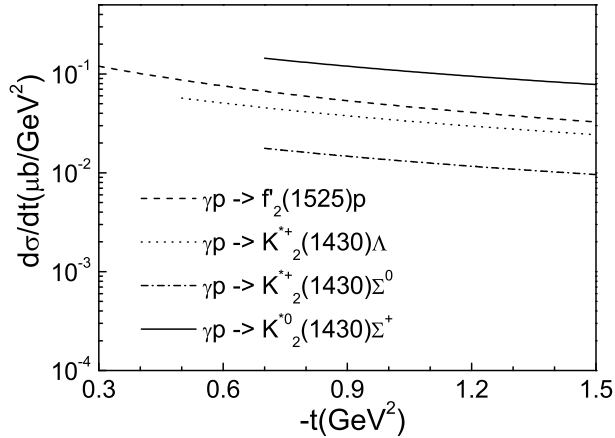}
\caption{Differential cross sections $\frac{d\sigma}{dt}$ as
functions of $t$ at $E_{\gamma} = 3.4$ GeV.} \label{Fig:dsigdt}
\end{center}
\end{figure}

Fig.~\ref{Fig:dsigdt} shows $d\sigma/dt$ at $E_{\gamma} = 3.4$
GeV for the four reaction modes studied. We see that
the slopes for the four reactions are quite similar.

\begin{figure}[htbp]
\begin{center}
\includegraphics[scale=0.45]{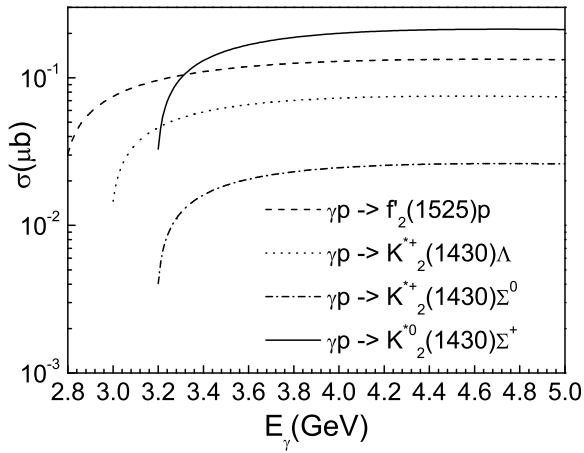}
\caption{Total cross sections for the $\gamma p \to f'_2(1525)p$ and
$\gamma p \to K^*_2(1430)\Lambda(\Sigma)$ reactions as functions of
$E_{\gamma}$.} \label{Fig:tcs}
\end{center}
\end{figure}

In addition to the differential cross section, we calculate also the
total cross section for the four reactions as a function of the
photon beam energy $E_{\gamma}$. The results are shown in Fig.~\ref{Fig:tcs}.
The cross sections increase rapidly away from threshold and soon becomes almost constant at
higher photon energies.

We stress again that the reaction formalism advocated here involves no
free parameters, which allow us to make predictions for total cross sections.
The differential and total cross sections can be checked in future experiments,
such as those at CLAS. In this sense, the reaction mechanism can be easily tested.

\section{Conclusions} \label{sec:summary}

In recent years, it has been found that the $f'_2(1525)$ and
$K^*_2(1430)$ resonances, though long been accepted as ordinary
$q\bar{q}$ states, can be dynamically generated from the vector
meson-vector meson interaction, and therefore qualify as
vector-vector molecules. Many studies adopting such a scenario have
been performed in mesonic reactions and all yield positive results.
In the present work, we have proposed to test the molecular picture
in the photonuclear reaction. The elements needed for the test are
very simple, which makes particularly transparent the interpretation
of the results. On one side the $f'_2(1525)$ and $K^*_2(1430)$
couple to $VV$ in $I=0$ and $I=1/2$, respectively, and the couplings
have been fixed before in the unitary approach that generates the
$f'_2(1525)$ and $K^*_2(1430)$ as a $VV$ molecule based on the local
hidden gauge formalism for the interaction of vector mesons. On the
other side, with these couplings and the vector meson dominance
hypothesis, incorporated in the local hidden gauge approach, the
photon gets converted into one of the vector mesons, which interact
with the vector meson emitted by the incoming proton to generate the
$f'_2(1525)$ and $K^*_2(1430)$ resonances. With this simple picture
we predict both the differential and total cross sections, which
could be tested by future experiments, such as those at CLAS.

\section*{Acknowledgments}

One of us, E. O., wishes to acknowledge support from the Chinese
Academy of Science (CAS) in the Program of Visiting Professorship
for Senior International Scientists (Grant No. 2013T2J0012). L.S.G. thanks the Institute for Nuclear Theory at University of Washington
for its hospitality and the Department of Energy for partial support during the completion of this work.
This work is partly supported by the Spanish Ministerio de Economia y
Competitividad and European FEDER funds under the contract number
FIS2011-28853-C02-01 and FIS2011-28853-C02-02, and the Generalitat
Valenciana in the program Prometeo II-2014/068. We acknowledge the
support of the European Community-Research Infrastructure
Integrating Activity Study of Strongly Interacting Matter (acronym
HadronPhysics3, Grant Agreement n. 283286) under the Seventh
Framework Programme of EU. This work is also partly supported by the
National Natural Science Foundation of China under Grant Nos.
11475227,1375024, and 11522539. This work is also supported by the
Open Project Program of State Key Laboratory of Theoretical Physics,
Institute of Theoretical Physics, Chinese Academy of Sciences, China
(No.Y5KF151CJ1).

\bibliographystyle{plain}

\end{document}